\documentclass[aps,10pt,prx,twocolumn,showkeys,reprint]{revtex4-2}
\usepackage{graphicx}
\usepackage{epstopdf}
\graphicspath{{./figs}}
\usepackage{dcolumn}
\usepackage{bm}
\usepackage{amsmath}
\usepackage{amsfonts}
\usepackage{xcolor}
\usepackage{hyperref}
\hypersetup{colorlinks=true,linkcolor=red,urlcolor=blue,citecolor=blue}
\begin{document}
\title{Process-tensor approach to full counting statistics of charge transport in quantum many-body circuits}
\keywords{Full Counting Statistics, Quantum Transport, Quantum Circuits, Process Tensors, Temporal Matrix Product States, Heisenberg XXZ Model.}
\author{Hari Kumar Yadalam}
\email{hari\_kumar.yadalam@kcl.ac.uk}
\affiliation{Department of Physics, King’s College London, Strand, London, WC2R 2LS, United Kingdom}
\affiliation{School of Physics, Trinity College Dublin, Dublin 2, Ireland}
\author{Mark T. Mitchison}
\email{mark.mitchison@kcl.ac.uk}
\affiliation{Department of Physics, King’s College London, Strand, London, WC2R 2LS, United Kingdom}
\affiliation{School of Physics, Trinity College Dublin, Dublin 2, Ireland}
\begin{abstract}
We introduce a numerical tensor-network method to compute the statistics of the charge transferred across an interface partitioning an interacting one-dimensional many-body lattice system with $U(1)$ symmetry. Our approach is based on a matrix-product state representation of the process tensor (also known as influence functional or influence matrix) describing the effect of the bulk system on the degrees of freedom at the interface, allowing us to evaluate a multi-time correlation function that yields the moment-generating function of charge transfer. We develop a scheme to truncate non-Markovian correlations which preserves the proper normalization of the process tensor and ensures the correct physical properties of the generating function. We benchmark our approach by simulating magnetization transport within the Heisenberg spin-$1/2$ XXZ brickwork circuit model at infinite temperature. Our results recover the correct transport exponent describing ballistic, superdiffusive, and diffusive transport in different regimes of the model. We also demonstrate anomalous transport encoded by a self-similar scaling form of the moment-generating function outside of the ballistic regime. In particular, we confirm the breakdown of Kardar-Parisi-Zhang universality in higher-order transport cumulants at the isotropic point. Our work paves the way for process-tensor descriptions of non-Markovian open quantum systems to address current fluctuations in strongly interacting systems far from equilibrium.

\end{abstract}
\maketitle

\section{Introduction}

Understanding the dynamics of excitations in quantum many-body systems is a central question in modern theoretical physics. The most straightforward experimental probe of these excitations is to measure how the system responds to external perturbations, such as currents induced by external field gradients~\cite{Forster2019Hydrodynamic,Mazenko2006Nonequilibrium}. In the well-studied linear-response regime, fluxes are proportional to the applied field, with coefficients determined by current-current correlation functions that encode information on a subclass of excitations in the system~\cite{Kubo1957Statistical1,Kubo1957Statistical2,Groot2013Nonequilibrium}. These transport coefficients can be used to identify distinct universality classes, e.g.~the low-frequency behavior of the conductivity determines whether transport is subdiffusive, diffusive, superdiffusive, or ballistic. 

Transport phenomena also open a vital window on universal physics far from equilibrium. In one-dimensional (1D) spin chains with nearest-neighbor interactions, a panoply of different high-temperature transport properties have been found within and beyond linear response, depending on symmetries, integrability, and nonequilibrium boundary conditions~\cite{Bertini2021Finite,Landi2022Nonequilibrium}. These discoveries have been driven, on the one hand, by experiments on controlled quantum many-body systems~\cite{Hild2014UCT,Jepsen2020ST,Scheie2021KPZ,Keenan2023KPZ} and, on the other hand, by novel theoretical developments including tensor-network methods~\cite{prosenMatrixProductSimulations2009, Prosen2011Open,Prosen2011Exact, rakovszkyDissipationassistedOperatorEvolution2022, Yao2025Hidden, Popkov2025Exact} and the framework of generalized hydrodynamics~\cite{olalla2016ghd,Bertini2016Transport,Doyon2020Lecture,Essler2023Short,Doyon2025Generalized,doyon2025nonlinear}. The rise of digital quantum simulation has also sparked new interest in understanding transport within quantum circuits evolving in discrete time, which may either replicate~\cite{Marko2019Ballistic} or radically depart~\cite{Zoller2019Trotter1,Zoller2019Trotter2,Chinni2022Trotter,Piroli2023Trotter1,Summer2024Anomalous,Piroli2025Trotter2,Zoller2025Trotter3}  from their continuous-time counterparts.

However, recent results have demonstrated that average currents are insufficient to characterize universality classes of high-temperature transport in 1D~\cite{Gopalakrishnan2023Anomalous,Gopalakrishnan2024Distinct,Rosenberg2024Dynamics,Valli2025Efficient}.
Focus has therefore shifted to the fluctuations encoded in the full counting statistics (FCS) of conserved charges. While these fluctuations have long been studied in quantum optics and mesoscopic electron transport~\cite{Levitov1993Charge,Levitov1996Electron,Esposito2009Nonequilibrium,Nazarov2012Quantum,Landi2024Current}, FCS can now be probed directly in isolated quantum many-body lattice systems via site-resolved measurements~\cite{Wei2022Quantum,Wienand2024Emergence,Rosenberg2024Dynamics,Joshi2025Measuring}. In this context, traditional weak-coupling approximations do not apply and thus predicting the FCS in quantum many-body systems remains a challenging theoretical problem, not least because FCS is defined by a multi-time measurement, i.e., it is not a standard observable. Important analytical results have recently been obtained in particular limits of strong interactions or large spatiotemporal scales~\cite{Doyon2020BFT, Bertini2023NonequilibriumFCS, McCulloch2023Full, Gopalakrishnan2024Distinct, Fujimoto2026Exact, Yoshimura2026Anomalous}, but numerical methods remain indispensable for analyzing the microscopic physics from first principles~\cite{Gopalakrishnan2024Distinct,Samajdar2024Turnstiles,Valli2025Efficient,Andrew2026HardRod}.

In this work, we introduce a general tensor-network algorithm for computing the fluctuating charge transferred across an interface partitioning a 1D quantum many-body lattice with $U(1)$ symmetry. By interpreting the interface as an open quantum system, we recast FCS as the contraction of so-called process tensors representing the influence of the environments on either side of the interface. The process tensor---the most general description of local, multi-time operations on a quantum system~\cite{Pollock2018Non,Milz2021Non}---underpins numerous tensor-network algorithms to simulate non-Markovian dissipation~\cite{Strathearn2018TEMPO, Jorgensen2019Exploiting, Cygorek2022ACE, Thoenniss2023FTEMPO1,Thoenniss2023FTEMPO,Link2024Open,Keeling2025Process,Sonner2025Semigroup} and local observables in isolated lattice models~\cite{Banuls2009TMPS,Hastings2015TE, Lerose2021IM,Sonner2021IM,Carignano2025TMPS}. Here, we adopt the light-cone folding algorithm~\cite{Miguel2022Light,Lerose2023Overcoming} to efficiently represent each process tensor as a temporal matrix product state (MPS). The bond dimension of this MPS quantifies environmental memory~\cite{Cygorek2025Inner}, which is expected to grow slowly with time for integrable systems~\cite{Lerose2021TMPS, Giudice2022TMPS,Foligno2023GF, Carignano2024TMPS, Wang2025TMPS} and also some chaotic systems under appropriate coarse-graining~\cite{Peter2025TMPS,Vilkoviskiy2025Temporal}. To capture these memory effects efficiently at long timescales, we develop an original MPS compression scheme that preserves the proper normalization of each process tensor, which may be of independent interest, e.g., for boundary impurity problems with a single environment.

To showcase our approach, we study magnetisation transport across a discrete-time quantum circuit version of the spin-$1/2$ XXZ model at infinite temperature, motivated by recent experiments~\cite{Wei2022Quantum,Rosenberg2024Dynamics}. Analogous to its continuous-time counterpart~\cite{Bertini2016Transport}, the XXZ circuit exhibits ballistic, superdiffusive, and diffusive transport depending on the anisotropy parameter, $\Delta$~\cite{Marko2019Ballistic}. Of particular interest is the isotropic point $\Delta = 1$, where the conductance is superdiffusive~\cite{Ljubotina2017KPZ, Scheie2021KPZ, Wei2022Quantum, Keenan2023KPZ}, consistent with the celebrated Kardar-Parisi-Zhang (KPZ) universality class~\cite{Kardar1986KPZ, Marko2019KPZ}. Our results show that KPZ universality breaks down for infinite-temperature cumulants above second order, corroborating recent experimental~\cite{Rosenberg2024Dynamics} and theoretical~\cite{Valli2025Efficient} findings. We also find ballistic transport with approximately Gaussian fluctuations for $|\Delta| <1$ and diffusive transport with highly non-Gaussian fluctuations for $|\Delta|>1$. Outside of the ballistic regime, we show that the moment-generating function adopts a self-similar scaling form which implies the breakdown of the large-deviation principle, signalling that both diffusion and superdiffusion in this model are highly anomalous.

We note that Ref.~\cite{Valli2025Efficient} recently introduced a complementary numerical method to compute FCS in many-body lattices and applied it to the infinite-temperature XXZ model. In Ref.~\cite{Valli2025Efficient}, however, the time-dependent cumulant generating function is represented in terms of a matrix-product operator (MPO) that is evolved by conventional tensor-network time propagation methods. The accuracy of this MPO representation is limited by operator entanglement~\cite{Dowling2026Classical}, which is expected to grow at most logarithmically in integrable systems and linearly in nonintegrable ones~\cite{Prosen2007LOE}. By contrast, the efficiency of our temporal MPS representation is limited by the growth of so-called temporal entanglement. While early evidence suggests that temporal entanglement should scale similarly to local operator entanglement in a given model~\cite{Lerose2021TMPS, Giudice2022TMPS, Foligno2023GF, Carignano2024TMPS, Jiangtian2024Temporal}, recent work has found slow temporal entanglement growth in \textit{nonintegrable} systems under certain conditions~\cite{Wang2024Exact, Wang2025TMPS, Vilkoviskiy2025Temporal}, suggesting that our approach could be applied to study fluctuating charge transport beyond integrable parameter regimes. We also note two previous works that have used process tensors to numerically compute the FCS of dissipative energy transfer, albeit in the different context of open quantum systems coupled to noninteracting environments~\cite{Popovic2021FCS, Shubrook2025FCS}. By establishing that a similar philosophy is fruitful in the many-body context, we hope that our work will inspire further applications of process-tensor methods in the nonequilibrium thermodynamics of strongly correlated systems.

We begin by introducing the general problem of FCS in quantum circuits in Sec.~\ref{sec:model}, before describing our process-tensor methodology and MPS compression scheme in Sec.~\ref{sec:methodology}. Results for the XXZ circuit model are presented in Sec.~\ref{sec:results_XXZ}, where we compute the evolution of a local magnetisation correlator as well as the FCS, computing charge cumulants up to sixth order for different values of the anisotropy parameter and timestep. We provide overall conclusions and an outlook on future work in Sec.~\ref{sec:conclusion}. Numerical convergence analysis is delegated to the appendix.

\section{Quantum Circuit model and Full Counting Statistics}
\label{sec:model}

We consider an infinite 1D lattice of qudits evolving under a unitary brickwork circuit with $U(1)$ symmetry. Brickwork circuits are fundamental models of nonequilibrium many-body physics with local interactions, which can also be interpreted as a Trotter-Suzuki approximation of continuous-time Hamiltonian dynamics with nearest-neighbour couplings~\cite{Bertini2025Exactly,Bertini2026Circuits}. The Hilbert space of the model system is $\mathbb{H}=\bigotimes_{l\in \mathbb{Z}} \mathbb{H}_l$, where each of the local spaces $ \mathbb{H}_l$ is isomorphic to $\mathbb{C}^d$, with $d$ the local qudit dimension. The evolution operator for a single time step is given by 
\begin{eqnarray}
\mathbb{U}&=& \mathbb{U}_{{\rm e}}\mathbb{U}_{{\rm o}},
\end{eqnarray}
where $\mathbb{U}_{{\rm e}/{\rm o}}$ correspond to unitary gates that act on all even/odd bonds of the lattice, being given by
\begin{eqnarray}
    \mathbb{U}_{{\rm e}} &=& \prod_{l\in \mathbb{Z}} \mathbb{U}_{2l,2l+1}\nonumber\\
     \mathbb{U}_{{\rm o}} &=& \prod_{l\in \mathbb{Z}}^{} \mathbb{U}_{2l-1,2l}.
\end{eqnarray}
The two-site qudit gates $\mathbb{U}_{l,l+1}$ act non-trivially on the two site Hilbert space $\mathbb{H}_l\bigotimes\mathbb{H}_{l+1}$. We assume that the two site gates and hence the circuit commute with a local conserved charge
\begin{eqnarray}
    \mathbb{M}&=& \sum_{l\in \mathbb{Z}}^{} \mathbb{M}_l,
\end{eqnarray}
which generates the $U(1)$ symmetry. 

We are interested in studying the transport of the conserved charge across a (fictitious) interface at site $j=0$, which separates two semi-infinite halves of the lattice to the left ($j\leq 0$) and right ($j>0$). To quantify this, we consider the probability that a quantity of charge $q$ is transferred from right to left after $n$ timesteps, $\mathcal{P}(q, n)$. This probability is defined by a two-point measurement protocol~\cite{Tasaki2000Jarzynski,Kurchan2000Quantum,Esposito2009Nonequilibrium,Campisi2011Colloquium}, wherein the charge difference operator $\Delta \mathbb{M}=\sum_{l\leq 0}^{}\mathbb{M}_l - \sum_{l>0}^{}\mathbb{M}_l$ is projectively measured at the initial and final times, yielding eigenvalues $\delta m_i$ and $\delta m_f$ as outcomes, respectively. The charge transported from right to left is $q = \delta m_f - \delta m_i$, which is distributed according to
\begin{eqnarray}
\label{prob_TPM}
    \mathcal{P}(q,n) &=& \sum_{\delta m_f , \delta m_i}^{}\delta( q -(\delta m_f-\delta m_i))\times\nonumber\\
    &&{\rm Tr}\left[\Pi_{\delta m_f}\mathbb{U}^n \Pi_{\delta m_i}\rho(0)\Pi_{\delta m_i} {\mathbb{U}^\dag}^n\right],
\end{eqnarray}
where $\Pi_{\delta m}$ is the projection operator into the eigenspace of $\Delta \mathbb{M}$ with eigenvalue $\delta m$ and $\rho(0)$ is the initial state. 

For computations, it is convenient to work with the Fourier transform of $P(q,n)$: the moment-generating function 
\begin{equation}
    \label{MGF_def}
    \mathcal{Z}(\lambda,n)= \int_{-\infty}^\infty dq \, e^{i \lambda q}\mathcal{P}(q,n),
\end{equation}
where $\lambda$ is called the counting field. The cumulants of $q$, denoted $\kappa_r$,  are generated according to 
\begin{equation}
    \label{cumulants}
    \kappa_r(n) = (-i)^r \left.\frac{\partial^r}{\partial \lambda^r} \ln \mathcal{Z}(\lambda,n)\right\vert_{\lambda = 0},
\end{equation}
e.g., $\kappa_1 = \langle q\rangle$ is the mean charge transfer, $\kappa_2 = \langle q^2\rangle - \langle q\rangle^2$ is the variance, and so on. Assuming that $[\rho(0),\Delta \mathbb{M}]=0$, so the initial measurement does not perturb the state, Eqs.~\eqref{prob_TPM} and~\eqref{MGF_def} yield
\begin{eqnarray}
\label{MGF_vectorised}
    \mathcal{Z}(\lambda,n) &=& {\rm Tr}\left[e_{}^{i \lambda \Delta \mathbb{M}} \mathbb{U}^n e_{}^{-i \lambda \Delta \mathbb{M}}\rho(0) {\mathbb{U}^\dag}^n\right].
\end{eqnarray}
For concreteness, we consider an infinite-temperature initial state, $\rho(0) =\bigotimes_{n \in \mathbb{Z}}\frac{\mathbb{I}}{d}$, where $\mathbb{I}$ is the identity operator. 

It is convenient to work in a vectorized or folded representation using the map $|k\rangle \langle l| \to |l\rangle^* |k\rangle$, where the asterisk denotes complex conjugation in the computational basis~\cite{Landi2024Current}. The vectorized moment-generating function for an infinite-temperature state is then given by
\begin{eqnarray}
    \mathcal{Z}(\lambda,n) &=& (\mathcal{I}| e_{}^{i \frac{\lambda}{2}\Delta \mathcal{M}}\mathcal{U}^n e_{}^{-i \frac{\lambda}{2} \Delta \mathcal{M}}|\mathcal{I}),
\end{eqnarray}
where $|\mathcal{I})=\bigotimes_{j\in \mathbb{Z}}^{} |\mathcal{I}_1)$ is a product of maximally entangled pairs $|\mathcal{I}_1)=\frac{1}{\sqrt{d}}\sum_{k=0}^{d-1}|k\rangle |k\rangle$, i.e., the vectorized local infinite temperature state at each site. Moreover, $\mathcal{U} = \mathbb{U}^*\otimes\mathbb{U} $ is the vectorized representation of the unitary superoperator $\mathbb{U} \star \mathbb{U}^\dag$, while $\Delta\mathcal{M} = \Delta \mathbb{M}^* \otimes \mathbb{I} +  \mathbb{I}\otimes \Delta \mathbb{M}$ is the vectorized superoperator $\{\Delta \mathbb{M},\star\}$. 

Now, note that the charge difference $\Delta \mathbb{M}$ only changes when a charge moves across the left-right interface. This fact is reflected by the commutation relation $\left[\Delta \mathbb{M},\mathbb{U}_{j,j+1}\right] = \left[\mathbb{M}_0 - \mathbb{M}_1,\mathbb{U}_{j,j+1}\right] \propto \delta_{j0}$. Exploiting this relation in Eq.~\eqref{MGF_vectorised} yields
\begin{eqnarray}
\label{MGF_simplified}
    \mathcal{Z}(\lambda,n) &=& (\mathcal{I}| \mathcal{U}_\lambda^n |\mathcal{I}),
\end{eqnarray}
where $\mathcal{U}_\lambda=\mathbb{U}_{+\lambda}^* \bigotimes \mathbb{U}_{-\lambda}$ with $\mathbb{U}_\lambda = \mathbb{U}_{e,\lambda} \mathbb{U}_{o}$, with
\begin{equation}
    \label{tilted_unitary}
    \mathbb{U}_{e,\lambda} =  e^{i\lambda(\mathbb{M}_0-\mathbb{M}_1)}\mathbb{U}_{0,1}e^{-i\lambda(\mathbb{M}_0-\mathbb{M}_1)}\prod_{j\neq 0} \mathbb{U}_{2j,2j+1} .
\end{equation}
That is, the counting field  $\lambda$ ``tilts'' the two-site unitary gate $\mathbb{U}_{0,1}$ coupling the two halves of the system, while all other gates are left unchanged. For the infinite lattice, Eq.~\eqref{MGF_simplified} can be represented diagrammatically~\cite{Bertini2025Exactly,Bertini2026Circuits} as
\begin{eqnarray}
    \hspace{-1cm} \mathcal{Z}(\lambda,n)&=&\vcenter{\hbox{\includegraphics[width=.67\linewidth]{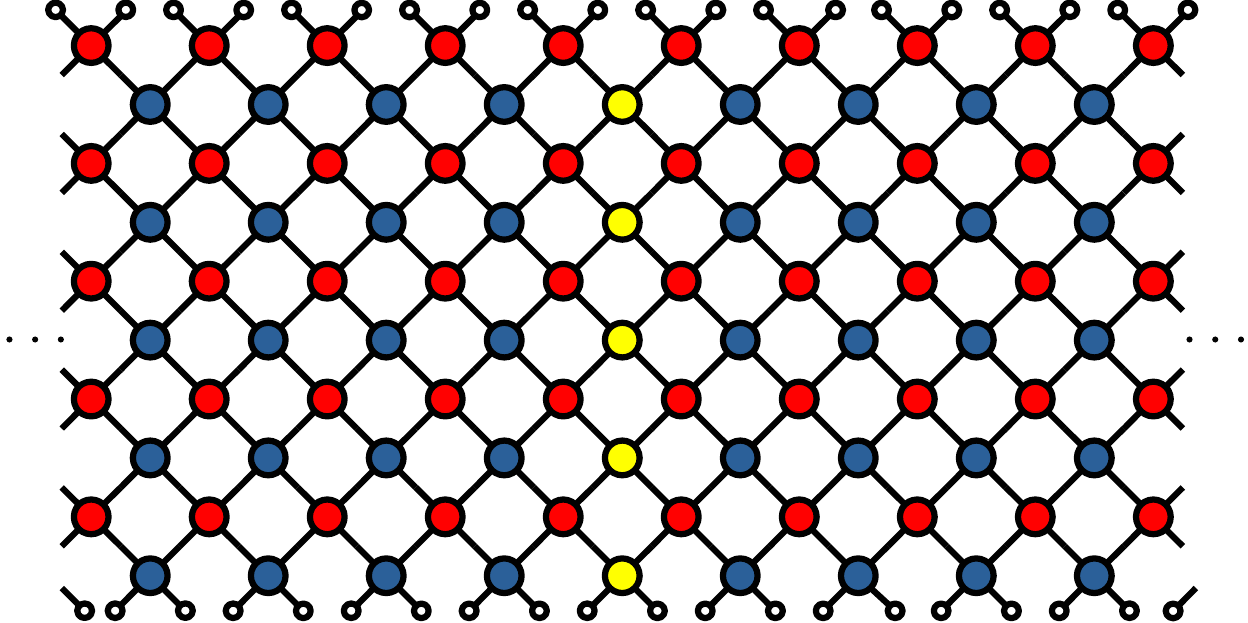}}},
\end{eqnarray}
with time increasing from bottom to top. Blue and red circles represent gates acting on odd-even and even-odd neighbouring sites, respectively. Small black pins represent the vectorized local infinite temperature density matrix at each site. Yellow circles represent the tilted unitary superoperator on sites $j=0,1$. 

Owing to the unitarity and unitality of the local two-site gates,
\begin{eqnarray}
\label{unitarity}
&&\vcenter{\hbox{\includegraphics[width=.08\linewidth]{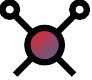}}}=\vcenter{\hbox{\includegraphics[width=.08\linewidth]{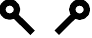}}} \hspace{.1\textwidth} \vcenter{\hbox{\includegraphics[width=.08\linewidth]{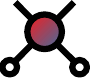}}}=\vcenter{\hbox{\includegraphics[width=.08\linewidth]{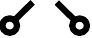}}}
\end{eqnarray}
an exact light cone structure emerges~\cite{Miguel2022Light,Lerose2023Overcoming,Bertini2025Exactly,Bertini2026Circuits}, which simplifies the network to
\begin{eqnarray}
\label{Z_lightcone}
    \mathcal{Z}(\lambda,n)&=&\vcenter{\hbox{\includegraphics[width=.33\linewidth]{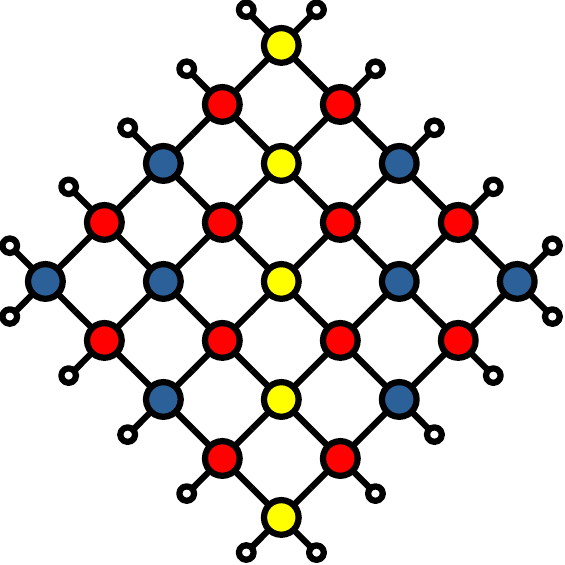}}}.
\end{eqnarray}
A similar diamond-shaped network arises for any initial product state of the form $\rho(0) = \bigotimes_{j\in \mathbb{Z}} \rho_j$ so long as the condition $[\mathbb{M}_j + \mathbb{M}_{j+1}, \rho_j\otimes \rho_{j+1} ] = 0$ is satisfied. 

\section{Computational methodology}
\label{sec:methodology}

\subsection{Computing FCS with temporal MPS}

Due to the exact light cone structure, the computation of the generating function is reduced to the contraction of the finite tensor network in Eq.~\eqref{Z_lightcone}. Nevertheless, exact contraction of this network has an exponential computational cost in the depth of the circuit, $n$. Conventional tensor-network algorithms such as time-evolving block decimation (TEBD)~\cite{Vidal2004Efficient,Verstraete2004MPDO,Schollwock2011MPS,Stoudenmire2010METS,Paeckel2019Time} would contract from bottom to top, i.e., sequentially in time, while approximating the intermediate state $\mathcal{U}_\lambda^m|\mathcal{I})$ generated after $m$ layers as an MPS with finite bond dimension and thus limited spatial correlations. This would reduce the complexity of a single evaluation of $\mathcal{Z}(\lambda,n)$ to quadratic in the circuit depth $n$. Nevertheless, a different network must be contracted for each value of $\lambda$ to compute the full statistics, which can be prohibitively expensive for large circuit depths.

Here we take a different approach, borrowing ideas from the temporal MPS literature~\cite{Banuls2009TMPS,Hastings2015TE,Lerose2021IM,Sonner2021IM,Miguel2022Light,Lerose2023Overcoming,Carignano2025TMPS,Keeling2025Process}. We interpret the central two sites as an open quantum system and construct tensors representing the influence of its environment to the left and right. That is, the computation of the generating function is interpreted as follows:
\begin{eqnarray}
\mathcal{Z}(\lambda,n)&=&\vcenter{\hbox{\includegraphics[width=.5\linewidth]{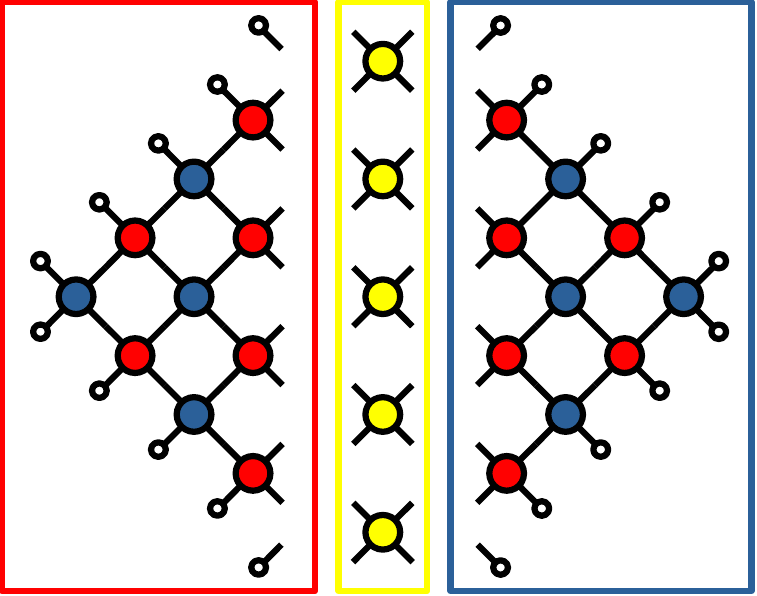}}}.
\end{eqnarray}
The tensors in the red and blue boxes are variously dubbed influence functionals~\cite{Feynman1963IF}, influence matrices~\cite{Lerose2021IM}, or process tensors~\cite{Pollock2018Non}. By approximating each tensor as a finite bond dimension MPS, 
\begin{eqnarray}
\label{tMPS}
   && \vcenter{\hbox{\includegraphics[width=.2\linewidth]{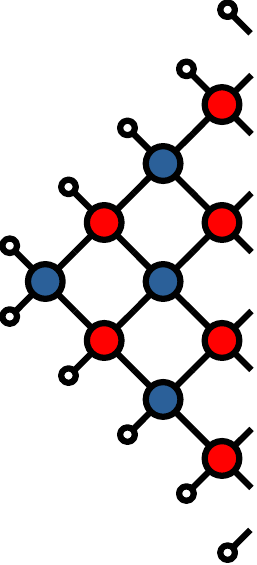}}}\approx\vcenter{\hbox{\includegraphics[width=.05\linewidth]{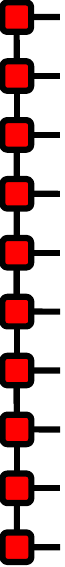}}}\hspace{.05\textwidth} \vcenter{\hbox{\includegraphics[width=.2\linewidth]{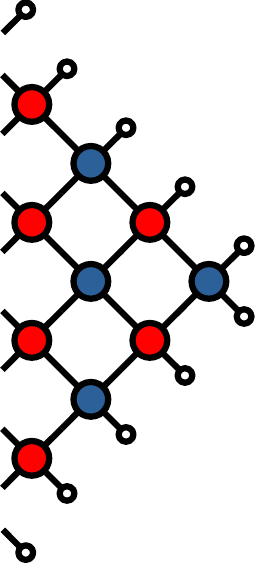}}}\approx \vcenter{\hbox{\includegraphics[width=.05\linewidth]{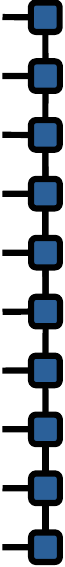}}}.
\end{eqnarray}
the computation of the generating function is rewritten as
\begin{eqnarray}
\label{Z_tMPS}
\mathcal{Z}(\lambda,n)&=&\vcenter{\hbox{\includegraphics[width=.12\linewidth]{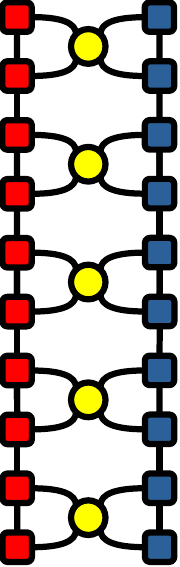}}}.
\end{eqnarray}
A key advantage of this approach is that the temporal MPSs are independent of the counting field and therefore need to be constructed only once for a given environment. Thus, the generating function $\mathcal{Z}(\lambda,n)$ at timestep $n$ is reconstructed by repeating the contraction in Eq.~\eqref{Z_tMPS} while varying $\lambda$ in the tilted unitary (yellow circles).

A similar construction can be used to compute other multi-time correlation functions, as shown already in Ref.~\cite{Banuls2009TMPS}. Consider the two-time correlation function at infinite temperature for a local observable $\mathbb{O}$ supported on site $j=0$:
\begin{equation}
    \langle \mathbb{O}(n) \mathbb{O}(0)\rangle = (\mathcal{I}| \mathcal{O} \mathcal{U}^n \mathcal{O}|\mathcal{I}),
\end{equation}
where $\mathcal{O} = \mathbb{O}^*\otimes \mathbb{I} $ is the superoperator describing multiplication by $\mathbb{O}$ from the left. In diagrammatic notation this can be expressed as
\begin{eqnarray}
\label{corr}
    \langle \mathbb{O}(n) \mathbb{O}(0)\rangle &=&\vcenter{\hbox{\includegraphics[width=.67\linewidth]{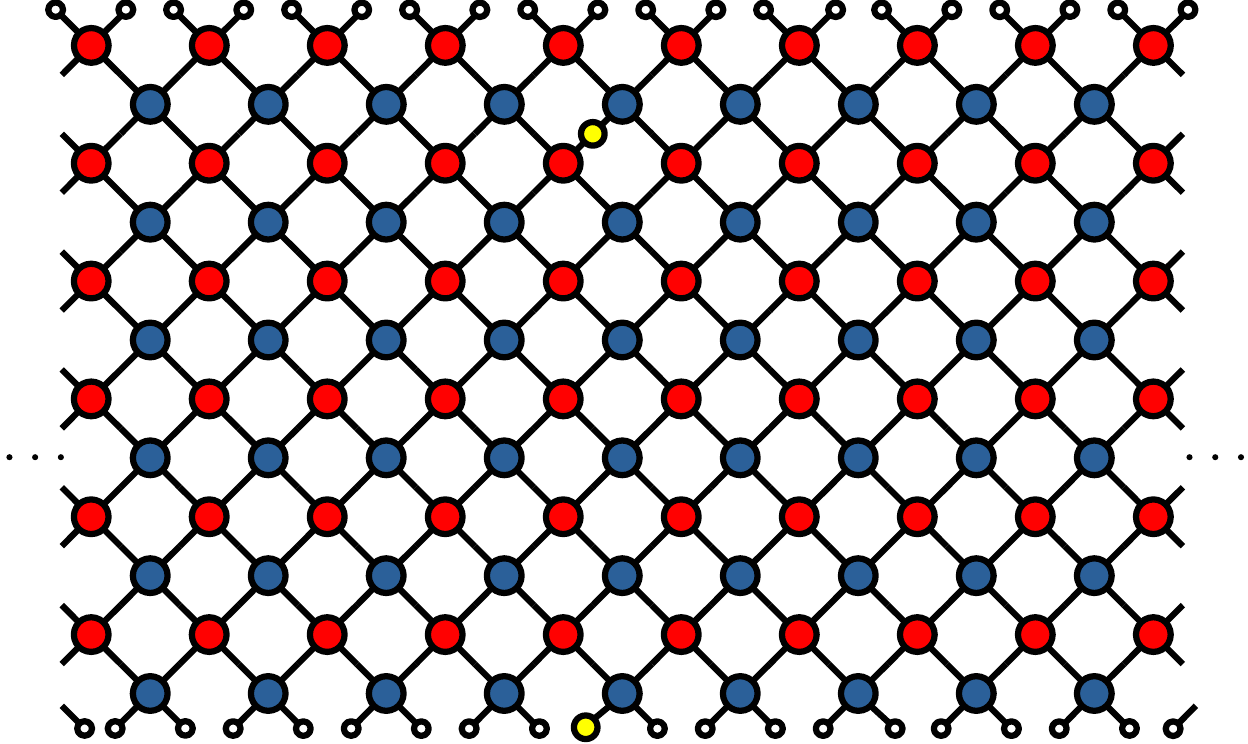}}}\nonumber\\
    &=&\vcenter{\hbox{\includegraphics[width=.41\linewidth]{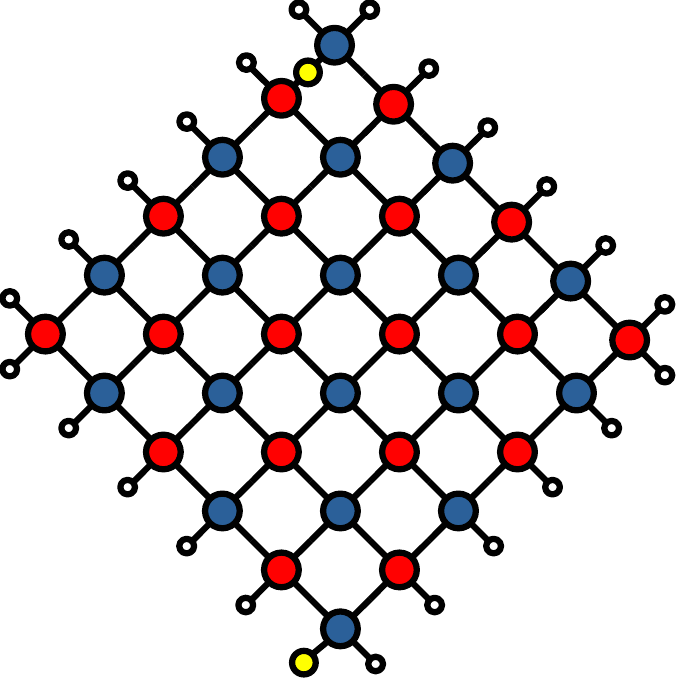}}} \approx \vcenter{\hbox{\includegraphics[width=.1\linewidth]{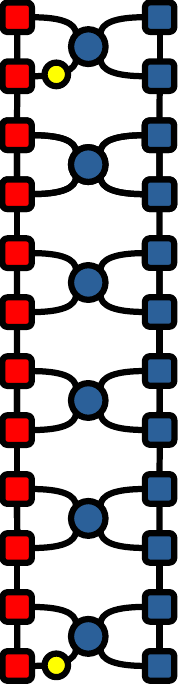}}},
\end{eqnarray}
where the small yellow circles indicate the superoperator $\mathcal{O}$ inserted into the circuit. On the second line of Eq.~\eqref{corr}, the first equality results from the exact light-cone structure while the second equality results from the MPS approximation. 

In fact, the same temporal MPS can describe any multi-time operation on the central subsystem. Indeed, the free horizontal legs of the MPS in Eq.~\eqref{tMPS} are of dimension $d^2$. There are two free legs for each  odd-even layer of the circuit, which together have dimension $d^4$ and thus span the space of superoperators acting on a single qudit at each timestep. For example, these superoperators could be elements of a quantum instrument, whose contraction with the temporal MPS yields the probability of obtaining a given sequence of measurement outcomes~\cite{Milz2021Non}: it is in precisely this sense that we refer to these objects as process tensors. The vertical bonds of the temporal MPS are of dimension $\chi$ and describe the propagation of information forward in time via the environment, i.e.~non-Markovianity or temporal entanglement. The bond dimension $\chi$ can also be interpreted as the effective dimension of the environment seen by the system~\cite{Pollock2018Non,Milz2021Non,Cygorek2025Inner,Carignano2025TMPS,Keeling2025Process}. Since the exact bond dimension grows exponentially with circuit depth, the accuracy of our MPS approximation scheme rests upon efficiently truncating $\chi$ to a tractable value, while retaining the most important information about the environmental influence, as we now elaborate.

\subsection{MPS truncation scheme}

The left and right process tensors individually obey the following normalization conditions:
\begin{eqnarray}
\label{no_intervention}
   && \vcenter{\hbox{\includegraphics[width=.05\linewidth]{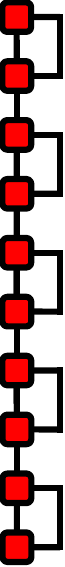}}}\approx\vcenter{\hbox{\includegraphics[width=.2\linewidth]{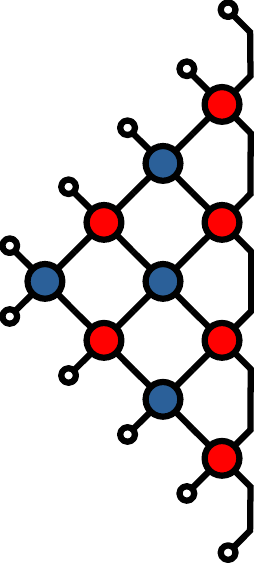}}}=1, \hspace{.05\textwidth} \vcenter{\hbox{\includegraphics[width=.05\linewidth]{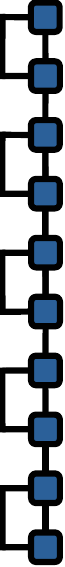}}}\approx\vcenter{\hbox{\includegraphics[width=.2\linewidth]{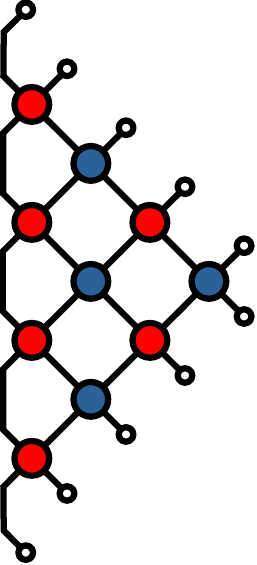}}}=1.\nonumber\\
\end{eqnarray}
At the level of the circuit, these conditions follow from trace preservation, e.g., as can be seen by applying Eq.~\eqref{unitarity} repeatedly to the right-hand side. At the level of the process tensor, this condition describes a ``no-intervention'' instrument, whose unique ``outcome'' (i.e., nothing) occurs with probability one. We therefore refer to this as the no-intervention normalization. Maintaining this normalization is important to preserve the physical properties of the generating function, e.g.~$\mathcal{Z}(0,n) = 1$. We note that alternate normalization conditions appropriate for computing the local correlator of a fixed operator have been explored in Refs.~\cite{Hastings2015TE,Miguel2022Light,Carignano2024TMPS}. 

We now present a numerical algorithm for computing temporal MPS representations while enforcing Eqs.~\eqref{no_intervention} at each step. Our algorithm draws its inspiration from the light-cone transverse folding approach~\cite{Banuls2009TMPS,Hastings2015TE,Lerose2021IM,Sonner2021IM,Miguel2022Light} and the density matrix truncation algorithm~\cite{White2018DMT}. Below, we present the construction of the left temporal matrix product state; the right state can be obtained similarly. The algorithm starts with the process for time step $n=1$,
\begin{eqnarray}
   && \vcenter{\hbox{\includegraphics[width=.03\linewidth]{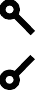}}}\quad, \nonumber
\end{eqnarray}
and iteratively constructs the MPS for the next time step by applying a spatial layer of gates to the MPS generated at the previous step. For example, passing from $n=4$ to $n=5$ we have
\begin{eqnarray}
\label{mps_iteration}
   && \vcenter{\hbox{\includegraphics[width=.08\linewidth]{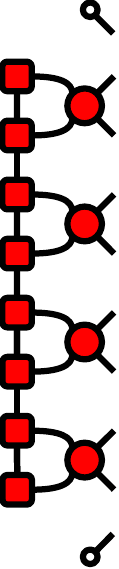}}}\,\approx \, \vcenter{\hbox{\includegraphics[width=.04\linewidth]{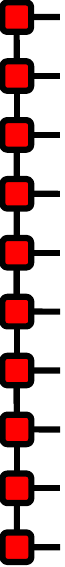}}}\quad,
\end{eqnarray}
where the right-hand side is an MPS compressed to a given bond dimension $\chi$. Standard MPS compression routines would either perform singular value decomposition (SVD) at each bond and discard the smallest singular values or use variational optimization to find the closest MPS with bond dimension $\chi$~\cite{Schollwock2011MPS}. However, this direct compression would generally violate the non-intervention normalization~\eqref{no_intervention}. 

Following Ref.~\cite{White2018DMT}, therefore, we first exploit the gauge freedom of the MPS representation to transform to a different basis, thus identifying those singular values that can be truncated without affecting the overall normalization. To describe this, we start from the no-intervention normalization condition after a single spatial layer of gates is applied:
\begin{eqnarray}
\label{no_intervention_overlap}
   && 1 = \vcenter{\hbox{\includegraphics[width=.08\linewidth]{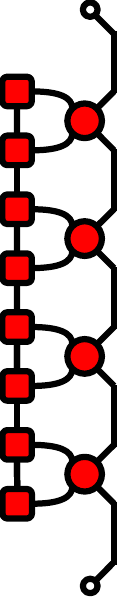}}}= \vcenter{\hbox{\includegraphics[width=.16\linewidth]{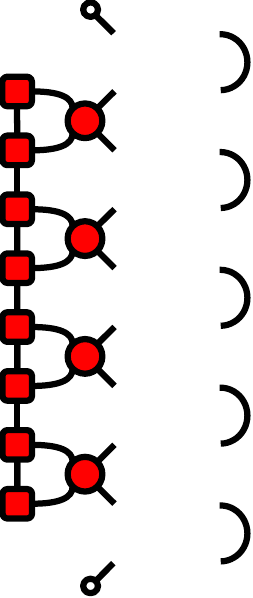}}}\,,
\end{eqnarray}
where the final equality expresses the normalization as the contraction of the tensor network on the left with a ``no-intervention'' MPS on the right. Now, consider truncation of the bond after applying the second gate from the top. We first apply the gate and do the exact SVD \textit{without} truncation:
\begin{eqnarray}
   && \vcenter{\hbox{\includegraphics[width=.1\linewidth]{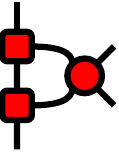}}} \overset{\text{SVD}}{=} \vcenter{\hbox{\includegraphics[width=.05\linewidth]{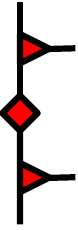}}}.
\end{eqnarray}
Inserting this into the normalization condition~\eqref{no_intervention_overlap} gives
\begin{eqnarray}
&& \vcenter{\hbox{\includegraphics[width=.2\linewidth]{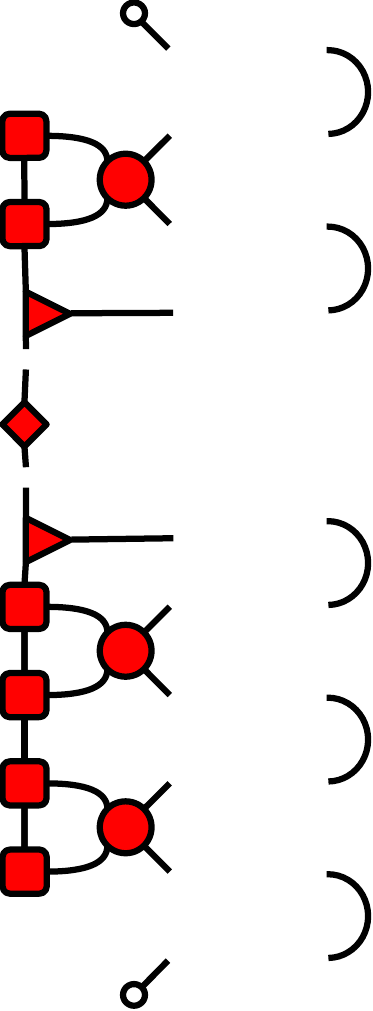}}} =1.
\end{eqnarray}
The dangling legs of the triangular tensors (above and below the diamond) index the basis states of the temporal bond. By an appropriate basis transformation, we single out a new basis vector that contributes to the contraction in Eq.~\eqref{no_intervention_overlap}, while all other orthogonal basis vectors do not. This is done by inserting unitaries and their inverses into the virtual bonds of the temporal matrix product state, as follows:
\begin{eqnarray}
\label{gauge_transformation}
&& \vcenter{\hbox{\includegraphics[width=.3\linewidth]{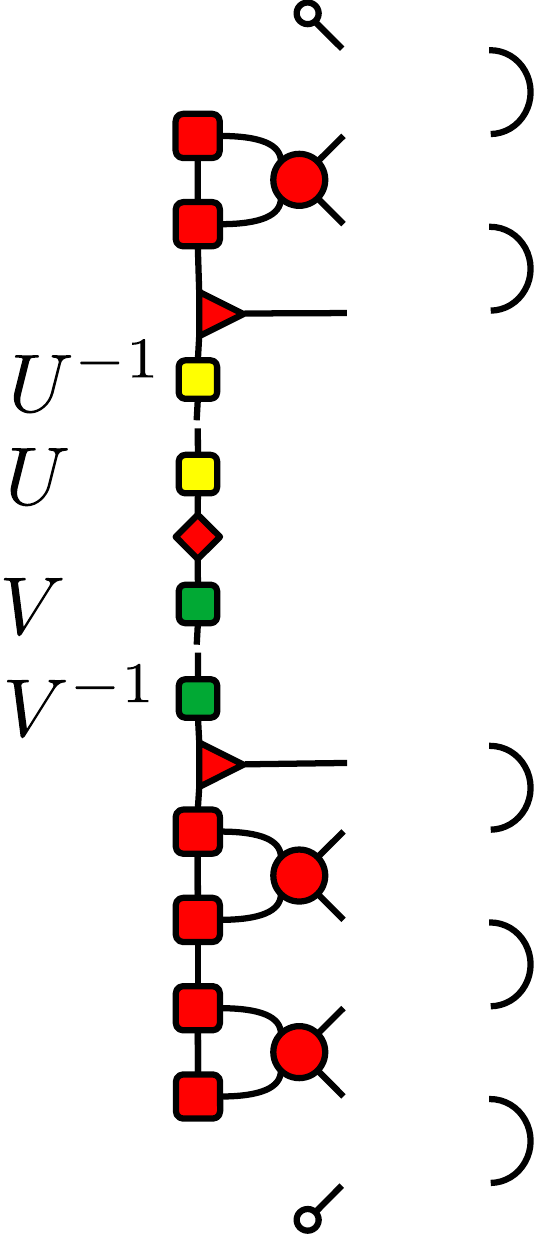}}} =1.
\end{eqnarray}
The unitaries $U$ and $V$ are defined by a QR decomposition of the ``temporal environments'' above and below the chosen bond. In this example, the future environment is decomposed as
\begin{eqnarray}
    \label{future_environment}
&& \vcenter{\hbox{\includegraphics[width=.12\linewidth]{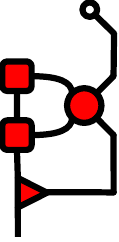}}}=\vcenter{\hbox{\includegraphics[width=.12\linewidth]{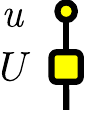}}}\,. 
\end{eqnarray}
As a consequence of the QR decomposition, the small yellow circle is a vector of the form $\mathit{u} = (u_1,0,0,\cdots)$ with only its first element nonzero, while the yellow square is a unitary matrix $U$. The past environment decomposes similarly as 
\begin{eqnarray}    \label{past_environment}
   && \vcenter{\hbox{\includegraphics[width=.12\linewidth]{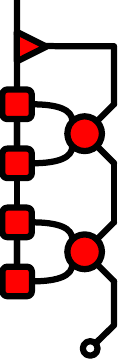}}}=\vcenter{\hbox{\includegraphics[width=.12\linewidth]{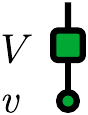}}}\,.
\end{eqnarray}
Inserting these into Eq.~\eqref{gauge_transformation} gives
\begin{eqnarray}
   && \vcenter{\hbox{\includegraphics[width=.12\linewidth]{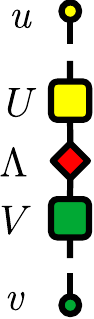}}}=1.
\end{eqnarray}
Since only the first elements of the vectors $\mathit{u}$ and $\mathit{v}$ are nonzero, only the $(1,1)$ matrix element of $U \Lambda V$ contributes to the no-intervention normalization. Hence, truncating the matrix $U \Lambda V=\begin{pmatrix} (1,1) & \\ & B \end{pmatrix}$ in the complementary block $B$ does not affect the normalization of the process tensor. This can be achieved by truncated singular value decomposition of $B$, i.e, $U\Lambda V\approx\begin{pmatrix} (1,1) & \\ & \bar{U}\bar{\Lambda}\bar{V} \end{pmatrix}$, as
\begin{eqnarray}
   && \vcenter{\hbox{\includegraphics[width=.04\linewidth]{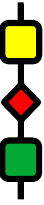}}} \overset{\text{approximate SVD of}  \,B}{\approx} \vcenter{\hbox{\includegraphics[width=.04\linewidth]{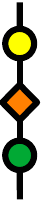}}}.
\end{eqnarray}
This decomposition yields a truncated MPS which exactly preserves normalization, since
\begin{eqnarray}
   && \vcenter{\hbox{\includegraphics[width=.2\linewidth]{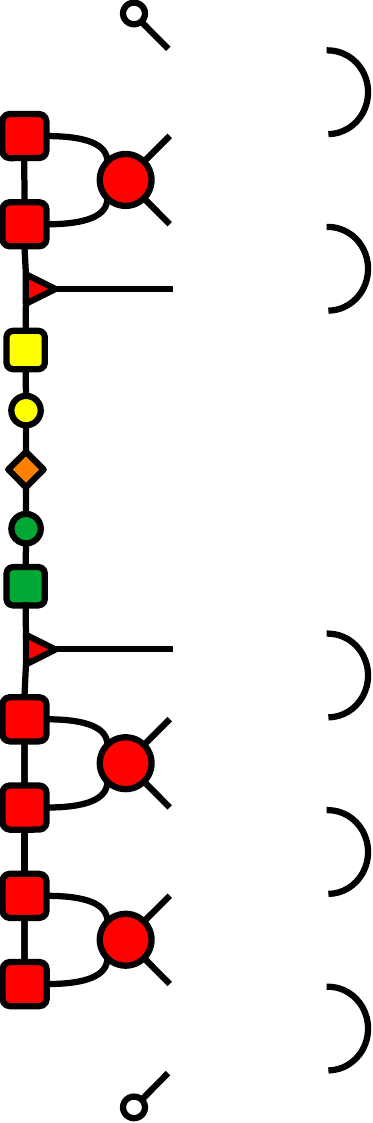}}} =1.
\end{eqnarray}
In summary, the two-site approximate gate application is
\begin{eqnarray}
   && \vcenter{\hbox{\includegraphics[width=.1\linewidth]{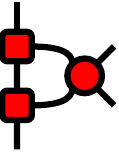}}} \approx \vcenter{\hbox{\includegraphics[width=.06\linewidth]{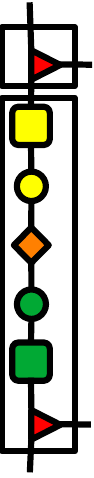}}} \equiv \vcenter{\hbox{\includegraphics[width=.05\linewidth]{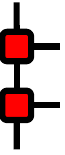}}},
\end{eqnarray}
where the last equality follows by contracting the tensors within the black boxes to form the updated local tensors of the MPS. A similar truncation is carried out at each two-site gate, sweeping from bottom to top, thus completing the MPS growth step shown in Eq.~\eqref{mps_iteration}. This growth is then iterated until the desired number of time steps $n$ is reached. Once MPS representations for the left and right process tensors are obtained, we contract with the tilted unitary to obtain the moment generating function as in Eq.~\eqref{Z_tMPS}.

\section{Application to magnetization transport in the XXZ circuit model}
\label{sec:results_XXZ}

\begin{figure*}[htbp!]
\centering
\includegraphics[width=\linewidth]{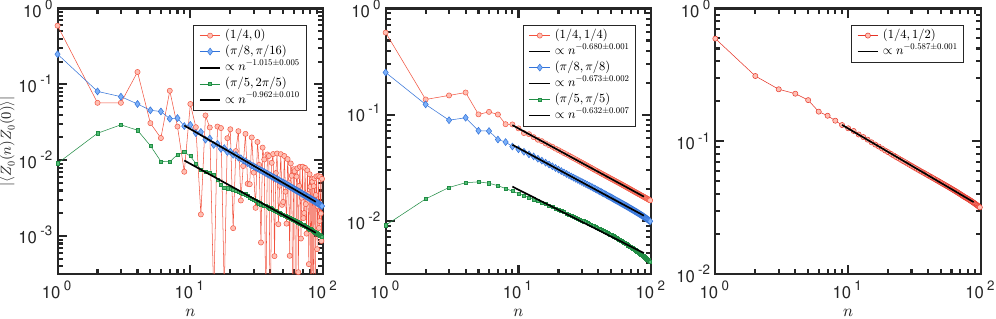}%
\caption{Local magnetization auto-correlator $\langle Z_0^{}(n) Z_0^{}(0)\rangle$ as function of circuit depth $n$ for various $(\mathcal{J},\mathcal{J}^\prime)$ parameters indicated in the legends that define the two-site gates of the spin-$\frac{1}{2}$ brickwork XXZ circuit model. Coloured lines with markers indicate the correlators obtained using the methodology presented in Sec.~\ref{sec:methodology}. Left, middle and right panels display the local correlators respectively in the ballistic, super-diffusive and diffusive transport regime of the circuit. Black lines indicate power-law fits to the data, and the respective exponents are displayed in the plot legends. Throughout this work, we use bond dimension $\chi=2^{10}$ in the ballistic regime (left panel) and $\chi=2^{11}$ in the superdiffusive and diffusive regime (middle and right panels, respectively).}
\label{fig:szsz}
\end{figure*}

We now apply our methodology to study magnetization transport in the brickwork spin-$\frac{1}{2}$ XXZ circuit model~\cite{Marko2019Ballistic,Rosenberg2024Dynamics,Valli2025Efficient}. The local dimension is $d=2$, with two-site gates given by
\begin{eqnarray}
    \mathbb{U}_{l,l+1} &=& e^{-i \mathcal{J}(X_{l}X_{l+1}+ Y_{l}Y_{l+1})-i\mathcal{J}_{}^\prime Z_l Z_{l+1}},
\end{eqnarray}
where $X_l, Y_l, Z_l$ are Pauli matrices at site $l$, and $\mathcal{J},\mathcal{J'}$ are coupling parameters. The conserved charge in this model is the total magnetization 
\begin{eqnarray}
   \mathbb{M} = \sum_l \mathbb{M}_l = \sum_l Z_l.
\end{eqnarray}
It is well established that the \textit{average} magnetization transport in this model shows qualitatively different behavior depending on the value of the anisotropy parameter, $\Delta = \mathcal{J}'/\mathcal{J}$, mirroring the behavior of the continuous-time model (obtained in the limit $\mathcal{J}\to 0$ for fixed $\Delta$)~\cite{Marko2019Ballistic}. At the free-fermion point, $\Delta = 0$, transport is ballistic as expected for a clean, noninteracting model. Magnetization transport remains ballistic for all positive values of $\Delta < 1$, becomes superdiffusive at the isotropic point $\Delta =1$, and is then diffusive for $\Delta > 1$. In the following, we use the XXZ circuit as a testbed to explore FCS in these different transport regimes with our method, recovering results in agreement with recent experimental~\cite{Rosenberg2024Dynamics} and numerical~\cite{Valli2025Efficient} investigations.

\subsection{Local magnetization correlator}

\begin{figure*}[htbp!]
\centering
\includegraphics[width=\linewidth]{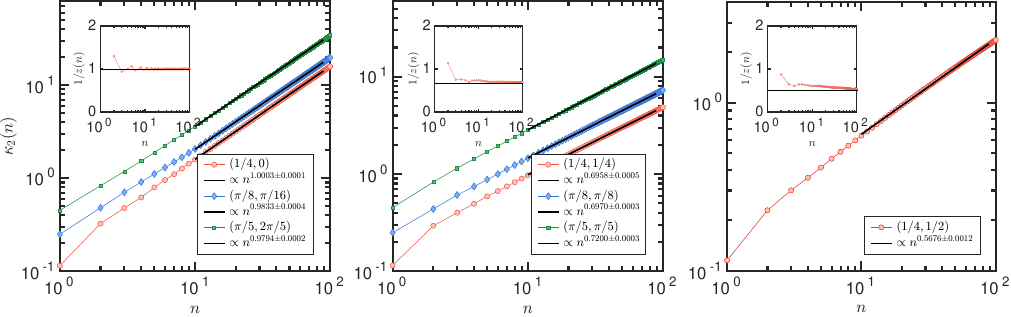}%
\caption{Second cumulant $\kappa_2(n)$ of the magnetization transported across the interface as a function of the circuit depth $n$ for the same choice of the parameters as in Fig.~\ref{fig:szsz}. Colored lines with markers indicate the numerical data obtained from the moment generating function as described in Sec.~\ref{sec:methodology} and the black lines indicate power-law fits to the data with the corresponding exponents indicated in the legends. Insets show the local time-dependent exponent for $\mathcal{J}=1/4$ [Eq.~\eqref{local_exponent}], which converges to the expected value (horizontal black line) for ballistic, superdiffusive, or diffusive transport in each of the three regimes.}
\label{fig:k2}
\end{figure*}

As an initial benchmark, we first compute the local magnetization correlator $\langle Z_0(n) Z_0(0)\rangle$. Such magnetization correlators have recently been probed experimentally in various digital quantum simulation platforms~\cite{Keenan2023KPZ,Shi2024QQS,Rosenberg2024Dynamics,Kumaran2025KPZ,Lee2026DQS}. Asymptotically, the local correlator is expected to decay as a power law with the circuit depth, $\langle Z_0(n) Z_0(0)\rangle \propto n^{-\frac{1}{z}}$, where the exponent $z$ identifies the transport universality class~\cite{Sirker2020Transport,Bertini2021Finite}. Specifically, $z = 1$ signifies ballistic transport and $z=2$ signifies diffusive transport, with intermediate values of $1 < z < 2$ corresponding to superdiffusive transport, while $z>2$ defines the subdiffusive regime. 

Fig.~\ref{fig:szsz} displays the infinite-temperature correlator computed using the methodology presented in Sec.~\ref{sec:methodology}, for various coupling parameters $(\mathcal{J}$, $\mathcal{J}')$ spanning the ballistic, superdiffusive, and diffusive regimes. One notable feature in the left panel of Fig.~\ref{fig:szsz} is the oscillatory behavior of the local correlator in the free-fermion limit where $\mathcal{J}' = 0$, which we attribute to the finite bandwidth induced by the discrete lattice. These oscillations are quickly damped in the presence of interactions between magnetic excitations ($\mathcal{J}'\neq 0$), and for finite values of $\mathcal{J}' < \mathcal{J}$ we observe the expected smooth algebraic decay of the correlator at long times, with ballistic exponent $z\approx 1$ obtained from a power-law fit of the data. At the isotropic point where $\mathcal{J}^\prime=\mathcal{J}$, the circuit model has enhanced $SU(2)$ symmetry and is expected to display super-diffusive transport with KPZ exponent $z=3/2$~\cite{Bulchandani2021Superdiffusion,Gopalakrishnan2023Anomalous,Gopalakrishnan2024Superdiffusion}. Our results for the isotropic point are presented in the middle panel of Fig.~\ref{fig:szsz}, and after a few transient circuit layers we observe approximate algebraic decay of the correlator. The fitted exponents are close to the KPZ value, with the  best agreement seen for smaller values of $\mathcal{J}$. Finally, the right panel of Fig.~\ref{fig:szsz} shows the local correlator in the diffusive regime, with $\mathcal{J}' = 2\mathcal{J} = 1/2$. We extract an exponent $z \approx 1.7$ from a fit of the data up to numerically accessible timescales. 

The exponents fitted from the correlator in Fig.~\ref{fig:szsz} show discrepancies from their expected values, especially in the diffusive regime and for larger $\mathcal{J}$. We attribute this primarily to the slow relaxation of the system, which prevents us from obtaining a reliable power-law fit on accessible simulation times. Naturally, we observe the slowest relaxation in the diffusive regime, which was also seen previously in Ref.~\cite{Valli2025Efficient}. As discussed in the next section, the FCS results in Fig.~\ref{fig:k2} provide a more accurate indicator of the transport exponent, showing that $z$ indeed converges to its expected values at late timescales. 

All results displayed in Fig.~\ref{fig:szsz} are well converged with respect to the bond dimension $\chi$, and we have also checked that our simulations reproduce exact results at the free-fermion and dual-unitary points. See the Appendix for detailed numerical benchmarking and convergence analysis. We find that convergence is most resource-intensive for larger values of $\mathcal{J}$ and in the diffusive or superdiffusive regimes, where bond dimensions of up to $\chi=2^{11}$ are needed to obtain convergence, whereas results in the ballistic regime are well converged for much smaller values of $\chi$. This suggests an intriguing connection between the efficiency of the temporal MPS representation and the hydrodynamic properties of the underlying model.

\subsection{FCS of magnetization transport}

\begin{figure*}[htbp!]
\centering
\includegraphics[width=\linewidth]{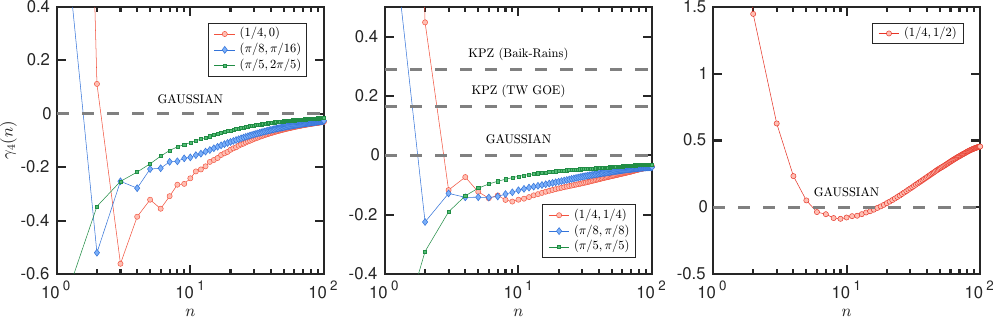}%
\caption{Kurtosis $\gamma_4 = \kappa_4/\kappa_2^2$ of the magnetization transported across the interface as a function of the circuit depth $n$, for the same parameters as in Figs.~\ref{fig:szsz} and \ref{fig:k2}. Horizontal dashed lines indicate predictions for the Gaussian distribution, i.e., $\gamma_4=0$. At the isotropic point (middle panel) we additionally display predictions for two non-Gaussian distributions in the KPZ universality class: the Baik-Rains and Tracy-Widom (TW) distributions.}
\label{fig:g4}
\end{figure*}

\begin{figure*}[htbp!]
\centering
\includegraphics[width=\linewidth]{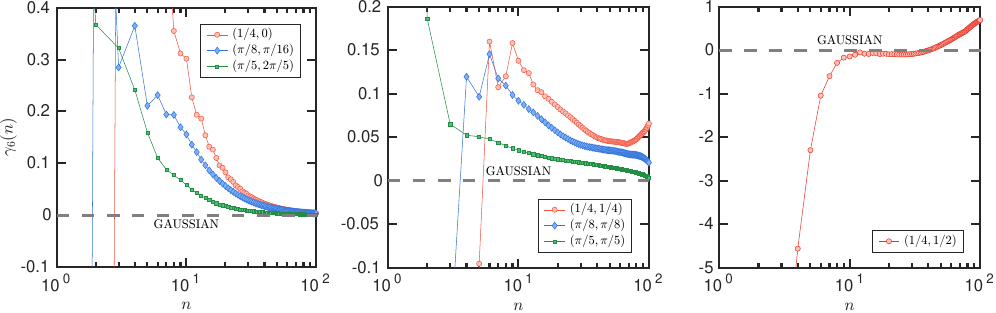}%
\caption{Sextosis $\gamma_6 = \kappa_6/\kappa_2^3$ of the magnetization transported across the interface as a function of the circuit depth $n$, for the same parameters as in Figs.~\ref{fig:szsz}--\ref{fig:g4}. The Gaussian sextosis of $\gamma_6=0$ is indicated by the horizontal dashed lines.}
\label{fig:g6}
\end{figure*}

To characterize the transported magnetization, we now present results for the time-dependent cumulants computed from Eq.~\eqref{cumulants} using the method described in Sec.~\ref{sec:methodology}. Due to the left-right reflection symmetry of the initial infinite-temperature state, the odd cumulants identically vanish. The first nontrivial cumulant is therefore the variance $\kappa_2$, while higher-order deviations from Gaussianity are captured by the kurtosis $\gamma_4 = \kappa_4/\left(\kappa_2\right)^{2}$ and sextosis $\gamma_6 = \kappa_6/\left(\kappa_2\right)^{3}$. 

Fig.~\ref{fig:k2} shows the variance as a function of circuit depth, with the same parameters and color scheme as in Fig.~\ref{fig:szsz}. We observe power-law growth of the variance in time, $\kappa_2(n) \propto n^{1/z}$. Fits to the data indicate that the transport exponent $z$ is close to its expected values for ballistic ($z=1$), superdiffusive ($1<z<2$) or diffusive ($z=2$) behavior. In all three regimes, the cleanest convergence of the transport exponent occurs for the smallest value of $\mathcal{J}=1/4$ that we consider. To make this convergence clear, the inset of Fig.~\ref{fig:k2} shows the local time-dependent exponent $z(n)$ defined by~\cite{Valli2025Efficient}
\begin{equation}
\label{local_exponent}
    \frac{1}{z(n)} = \frac{\Delta \log \kappa_2(n)}{\Delta \log n},
\end{equation}
with $\Delta f(n) = f(n+1)-f(n)$ the difference operator. In the right panel of Fig.~\ref{fig:k2} regime, we see that $z(n)$ continues to decrease over almost 100 circuit layers, as a consequence of the slow relaxation in the diffusive regime.

Next, we probe the Gaussianity of the FCS by plotting the kurtosis and sextosis of the distribution in Figs.~\ref{fig:g4} and~\ref{fig:g6}, respectively. Note that, while our results for the variance are well converged with bond dimension up to circuit depths of $n=100$, the results for kurtosis in the diffusive regime and the sextosis in both diffusive and superdiffusive regimes are sensitive to reductions in $\chi$ after $n\sim 64$ layers (we use a maximum bond dimension of $\chi=2^{11}$). See the Appendix for the detailed convergence analysis.

In the ballistic regime (left panels of Figs.~\ref{fig:g4} and ~\ref{fig:g6}), both the kurtosis and the sextosis converge to a value near zero at long times, consistent with asymptotically Gaussian fluctuations. At the isotropic point (middle panels of Figs.~\ref{fig:g4} and ~\ref{fig:g6}), the long-time kurtosis is small and negative, consistent with a weakly non-Gaussian asymptotic distribution. Notably, here the kurtosis remains far from the predictions of the Tracy-Widom~\cite{TracyWidom1994} or Baik-Rains~\cite{BaikRains2000} distributions within the KPZ universality class~\cite{Prahofer2004KPZ}. In the diffusive regime (right panels of Figs.~\ref{fig:g4} and ~\ref{fig:g6}), the kurtosis and sextosis continue to grow monotonically at the longest accessible times, indicating strongly non-Gaussian fluctuations~\cite{Gopalakrishnan2024Distinct}. We note that these results are fully consistent with those found via a different method in Ref.~\cite{Valli2025Efficient}, and in particular they confirm the breakdown of the KPZ and diffusive universality classes in higher-order FCS of the XXZ circuit model.

\begin{figure*}[htbp!]
\centering
\includegraphics[width=\linewidth]{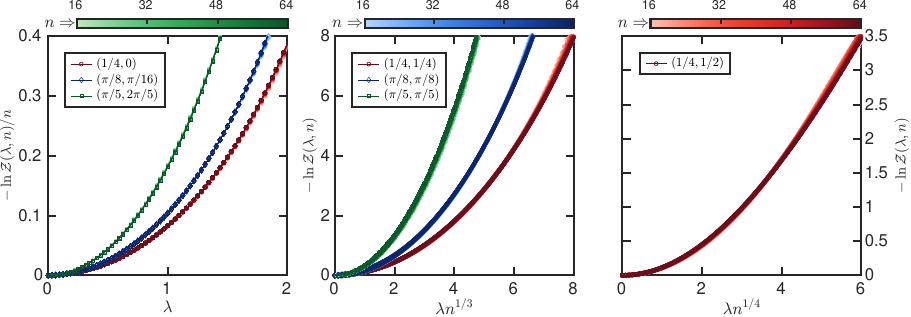}%
\caption{Cumulant-generating function $\ln\mathcal{Z}(\lambda,n)$ for the magnetization transported across the interface as a function of the counting field $\lambda$ for various circuit depths $n$ indicated in the colorbars. We use the same parameters and color scheme as all previous figures. In the ballistic transport regime (left panel), the cumulant generating function shows a large-deviation form, $\mathcal{Z}(\lambda,n) = e_{}^{-n \mathcal{F}(\lambda)}$, where the function $\mathcal{F}(\lambda)$ depends on the circuit parameters $(\mathcal{J},\mathcal{J}')$ but not on the circuit depth $n$. In the superdiffusive (middle panel) and diffusive (right panel) regimes, we see scaling collapse when plotting the cumulant generating function as a function of $\lambda n^{1/2z}$, indicating a self-similar form $\mathcal{Z}(\lambda,n)= \mathcal{Z}(\lambda n^{1/2z})$ with the exponents $z=3/2$ and  $z=2$, respectively. }
\label{fig:z}
\end{figure*}

Perhaps the clearest demonstration of this anomalous transport is given by Fig.~\ref{fig:z}, where we directly plot the cumulant-generating function. For $\mathcal{J}'< \mathcal{J}$, we see that $\ln \mathcal{Z}(\lambda,n)/n$ is independent of $n$, indicating the large-deviation form $\mathcal{Z}(\lambda,n) \sim e^{-n \mathcal{F}(\lambda)}$ as expected from the ballistic macroscopic fluctuation theory~\cite{Jason2020BFT,Doyon2020BFT,Vecchio2022BFT,doyon2023ballistic,doyon2023emergence} and from space-time duality arguments~\cite{Bertini2023NonequilibriumFCS}. However, for $\mathcal{J}'\geq \mathcal{J}$, we instead see scaling collapse to the functional form $\mathcal{Z}(\lambda,n) = \mathcal{Z}(\lambda n^{1/2z})$, with $z = 3/2$ at the isotropic point and $z = 2$ in the diffusive regime, consistent with previous results~\cite{Gopalakrishnan2024Distinct,Valli2025Efficient}. Referring to Eq.~\eqref{cumulants}, this scaling form implies that the asymptotic cumulants scale differently with time at each order, as $\kappa_r(n)\sim n^{r/2z}$. This further implies that the kurtosis and sextosis ratios should eventually tend to constants, although this would occur beyond the numerically accessible timescales shown in Figs.~\ref{fig:g4} and~\ref{fig:g6}. In the diffusive regime, in particular, evidence from the exact analytical~\cite{Krajnik2022Exact} and numerical~\cite{Krajnik2024Dynamical} computations on classical many-body models suggest the limiting values $\gamma_4\to \frac{3}{2}\left(\pi -2\right)$ and $\gamma_6 \to \frac{15}{2}\left(4-\pi\right)$ as $n\to \infty$. The same values are also predicted by a nested-Gaussian form for the asymptotic spin-current distribution derived recently~\cite{Yoshimura2026Anomalous}. The results in Figs.~\ref{fig:g4} and~\ref{fig:g6} indicate that this asymptotic behavior should emerge only after many hundreds of circuit layers for the relatively low anisotropy parameters considered here.

\section{Conclusions}
\label{sec:conclusion}

In this work, we presented an efficient computational methodology for simulating full counting statistics of conserved charge transport under quantum circuit dynamics. By focusing on the transport of charge across an interface, we interpret the dynamics at the interface as an open quantum system problem, thus allowing us to exploit an MPS representation of the process tensors that encode the environment's influence. Inspired by Ref.~\cite{White2018DMT}, we developed a novel truncation scheme to preserve the proper normalization of the process tensor, which differs from standard pure MPS truncation methods~\cite{Schollwock2011MPS} and is also distinct from other truncation schemes recently introduced for light-cone folding algorithms~\cite{Hastings2015TE,Miguel2022Light,Carignano2024TMPS}. As we have shown, extraction of the moment generating function requires evaluation of a genuinely multi-time correlation function at the interface. In this sense, the computation of FCS pushes the process-tensor representation to its limits, whereas the majority of previous applications have focused on computing few lower-order correlation functions~\cite{Keeling2025Process,Chin2025Spectroscopy,Strunz2026Spectroscopy}.

To benchmark the performance of our method, we applied it to study magnetization transport in the XXZ circuit model, across ballistic, superdiffusive, and diffusive regimes. We were able to reproduce the expected transport exponent $z$ describing the charge variance $\kappa_2$ in each case, with the cleanest convergence found for small values of $\mathcal{J}$. Outside of the ballistic regime, we also demonstrated the self-similar scaling form of the moment generating function, $\mathcal{Z}(\lambda, n) = \mathcal{Z}(\lambda n^{1/2z})$, confirming the anomalous nature of subballistic transport in this integrable circuit model.

These simulations show that our process-tensor approach to FCS is competitive with the state-of-the-art tensor-network methodology introduced recently in Ref.~\cite{Valli2025Efficient}, obtaining similar results and reaching timescales beyond those accessible through state-vector time evolution by classical computation or recent quantum-simulator experiments~\cite{Rosenberg2024Dynamics}. Unlike our method which is limited by the growth of temporal entanglement within the process tensor, the approach of Ref.~\cite{Valli2025Efficient} is limited by the growth of operator entanglement of a specific unitary operator: the symmetry transformation generated by the conserved charge. The efficiency of both methods in this case is not surprising: temporal entanglement and local operator entanglement are both expected to grow sublinearly for integrable systems, and any homogeneous brickwork qubit circuit with $U(1)$ symmetry is integrable~\cite{Znidaric2025Integrability}. More generally, scaling of local operator entanglement has recently been argued to be the key factor determining simulation complexity for Heisenberg-picture evolution~\cite{Dowling2026Classical} as well as for temporal MPS representations of local correlation functions~\cite{Carignano2024TMPS}. Nevertheless, much remains to be understood about the connection between temporal and operator entanglement and the fundamental limitations of the process-tensor approach to local dynamics in many-body systems. We are especially motivated in this regard by recently discovered exact solutions for the process tensor~\cite{Klobas2020Matrix,Vilkoviskiy2025Properties} even in nonintegrable models~\cite{Wang2024Exact,Wang2025TMPS, Li2025Efficient}, and the demonstration that non-Markovianity can be significantly reduced by temporal coarse-graining~\cite{Peter2025TMPS,Vilkoviskiy2025Temporal}.  

A particularly interesting question for further investigation, therefore, is whether a systematic coarse-graining of the process tensor, e.g., such as the one introduced in Ref.~\cite{Dowling2024Tree}, can be used to efficiently capture hydrodynamic transport properties at even longer times than are currently accessible by our approach. We note that our normalization-preserving truncation algorithm can be generalized to preserve other quantities as well, e.g., few-point temporal correlators, which may be useful to retain relevant hydrodynamic information at long times, analogous to the density-matrix truncation approach~\cite{White2018DMT}. Moreover, our method can be straightforwardly generalized to describe continuous-time dynamics within the Trotter-Suzuki approximation, dissipative processes within the bulk (either real or artificially introduced to improve efficiency~\cite{rakovszkyDissipationassistedOperatorEvolution2022}), and out-of-equilibrium initial conditions. Exploiting internal symmetries and time-translation invariance~\cite{Link2024Open,Sonner2025Semigroup} to improve numerical efficiency are also natural next steps.

\section*{Acknowledgements}
We thank Bruno Bertini, Juan P. Garrahan, Benjamin Doyon, Takato Yoshimura and Stephen R. Clark for helpful discussions and comments on the manuscript. We gratefully acknowledge the financial support of the Royal Society via the University Research Fellowship \texttt{URF-T-261001}.
We acknowledge the ToCQS group computing node (KESHA) and  King's Computational Research, Engineering and Technology Environment (CREATE) for computational resources. 

\section{Appendix}
\subsection{Free fermions}
At $\mathcal{J}^\prime=0$, the brickwork qubit XXZ circuit becomes a match gate circuit, with the two site qubit gates given by, $\mathbb{U}_{n n+1}^{}=e_{}^{- i\mathcal{J}\left(X_{n}^{}X_{n+1}^{} + Y_{n}^{}Y_{n+1}^{}\right)}$. Local correlator and moment generating function for this match gate circuit can be computed exactly in polynomial time in circuit depth, by using Jordan-Wigner transformation and resultant gaussianity of dynamics and initial state. It is also important to realize that, owing to exact light cone structure, both the local correlator and the generating function for a lattice of size $2 N$ gives same result as in the thermodynamic limit for circuit of depth upto $N$. For the match gate case, we can use the Jordan-Wigner transformation from qubit operators to fermion operators as,
\begin{eqnarray}
    X_l &=& e_{}^{i\pi\sum_{l^\prime = -N+1}^{l-1}}\left[c_l^\dagger + c_l\right]\nonumber\\
    Y_l &=& -i e_{}^{i\pi\sum_{l^\prime = -N+1}^{l-1}}\left[c_l^\dagger - c_l\right]\nonumber\\
    Z_l &=& 2 c_l^\dagger c_l -1,
\end{eqnarray}
where $c_l$ and ($c_l^\dagger$) are fermionic annihilation and creation operator at site $l \in \{-N+1,-N+2,\cdots,N,N\}$ for the lattice of length $2 N$. With this the circuit unitary for a one time step in fermion representation can be written as,
\begin{eqnarray}
    \mathbb{U} &=& \mathbb{U}_o \mathbb{U}_e\nonumber\\
    \mathbb{U}_e &=& \bigotimes_{n \in \text{even}} e_{}^{-2i\mathcal{J} \left(c_{n}^\dagger c_{n+1}^{}+c_{n+1}^\dagger c_{n}^{}\right)} \nonumber\\
    \mathbb{U}_o &=& \bigotimes_{n \in \text{odd}} e_{}^{-2i\mathcal{J} \left(c_{n}^\dagger c_{n+1}^{}+c_{n+1}^\dagger c_{n}^{}\right)}. \nonumber\\
\end{eqnarray}
The local $Z$ correlator can be evaluated as follows,
\begin{eqnarray}
    &&\langle Z_0(n) Z_0(0)\rangle\nonumber\\
    &=& \langle\left(2 c_0^\dagger (n) c_0^{}(n) -1\right)\left(2 c_0^\dagger (0) c_0^{}(0) -1\right) \rangle \nonumber\\
    &=& 4\langle c_0^\dagger (n) c_0^{}(n)c_0^\dagger (0) c_0^{}(0)\rangle -4 \langle c_0^\dagger (0) c_0^{}(0)\rangle +1\nonumber\\
    &=& 4\langle c_0^\dagger (n) c_0^{}(0)\rangle \langle c_0^{}(n) c_0^\dagger(0)\rangle+4\langle c_0^\dagger (0) c_0^{}(0)\rangle^2\nonumber\\
    &&-4 \langle c_0^\dagger (0) c_0^{}(0)\rangle + 1 \nonumber\\
    &=& 4\langle c_0^\dagger (n) c_0^{}(0)\rangle \langle c_0^{}(n) c_0^\dagger(0)\rangle,
\end{eqnarray}
where to arrive at the second equality, we used time translation invariance, and to arrive at third, Wick's theorem is invoked~\cite{Wick1950Evaluation,Kamenev2023Field}, and the last invokes the correlator in infinite temperature state $\langle c_m^\dagger(0) c_n^{}(0)\rangle=\frac{\delta_{m n}}{2}$. Using the identity,
\begin{eqnarray}
    e_{}^{i \sum_{l,l^\prime =1}^{L} \mathcal{O}_{l l^\prime} c_l^\dagger c_{l^\prime}^{}} c_m e_{}^{-i \sum_{l,l^\prime =1}^{L} \mathcal{O}_{l l^\prime} c_l^\dagger c_{l^\prime}^{}} &=&\sum_{n=1}^{L}\left[e_{}^{-i \mathcal{O}}\right]_{m n} c_n^{},\nonumber\\
\end{eqnarray}
and it's conjugate, along with the zero time correlator in the infinite temperature state gives,
\begin{eqnarray}
    \langle Z_0(n) Z_0(0)\rangle &=& |\left[\mathcal{U}^n\right]_{00}^{}|^2,
\end{eqnarray}
where $\mathcal{U}$ is a single particle evolution operator defined as,
\begin{eqnarray}
    \mathcal{U} &=& \mathcal{U}_e \mathcal{U}_o\nonumber\\
    \mathcal{U}_o &=& \bigoplus_{ n \in \text{odd}} e_{}^{-2i \mathcal{J}\left[|n\rangle \langle n+1| + |n+1\rangle \langle n|\right]}\nonumber\\
    \mathcal{U}_e &=& \bigoplus_{n \in \text{even}} e_{}^{-2i \mathcal{J}\left[|n\rangle \langle n+1| + |n+1\rangle \langle n|\right]},\nonumber\\
\end{eqnarray}
with $\{|l\rangle\equiv c_l^\dagger |\Phi\rangle : l \in \{-N+1,-N+2,\cdots,N\}\}$ spanning the single particle Hilbert space of fermions (with the vacuum state $|\Phi\rangle$).

Similarly, the moment generating function for this match gate circuit can be obtained by using the determinant formula~\cite{Klich2002Full},
\begin{eqnarray}
    &&{\rm Tr}\left[e_{}^{i \sum_{l,l^\prime =1}^{L} \mathcal{O}_{l l^\prime}^{(1)} c_l^\dagger c_{l^\prime}^{}}\cdots e_{}^{i \sum_{l,l^\prime =1}^{L} \mathcal{O}_{l l^\prime}^{(K)} c_l^\dagger c_{l^\prime}^{}}\right]\nonumber\\ 
    &=& {\rm det}\left[\mathcal{I}+e_{}^{i \mathcal{O}_{}^{(1)}}\cdots e_{}^{i \mathcal{O}_{}^{(K)}}\right],
\end{eqnarray}
the moment generating function can be rewritten as follows,
\begin{eqnarray}
    \mathcal{Z}(\lambda,n) &=& \frac{1}{2^{4N+2}} {\rm Tr}\left[e_{}^{i\frac{\lambda}{2}\mathbb{T}}\mathbb{U}^n e_{}^{-i\frac{\lambda}{2}\mathbb{T}}\mathbb{U}^{-n}\right]\nonumber\\
    &=& {\rm det}\left[\frac{1}{2}\left(\mathcal{I}+e_{}^{i\frac{\lambda}{2}\mathcal{T}}\mathcal{U}^n e_{}^{-i\frac{\lambda}{2}\mathcal{T}} \mathcal{U}^{-n}\right)\right],
\end{eqnarray}
where $\mathbb{T}=\left[\sum_{l=-N+1}^{0}c_l^\dagger c_l^{}-\sum_{l=1}^{N}c_l^\dagger c_l^{}\right]$ and $\mathcal{T}=\sum_{l=-N+1}^{0} |l\rangle \langle l| -\sum_{l=1}^{N}|l\rangle \langle l|$.
\begin{figure*}[htbp!]
\centering
\includegraphics[width=\linewidth]{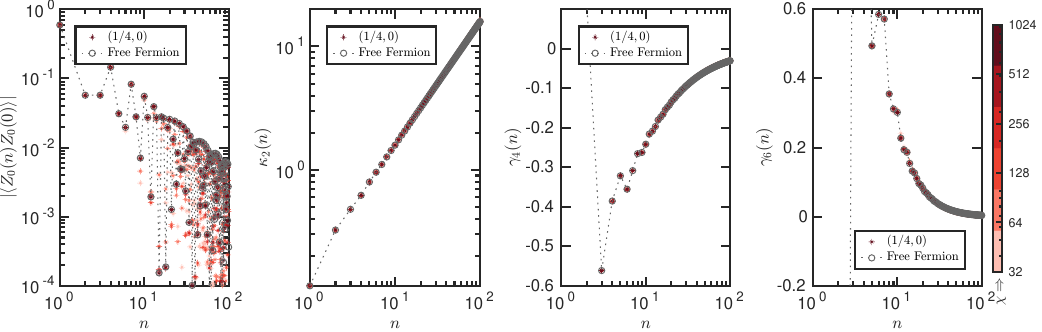}%
\caption{Benchmarks for the local correlator (first panel), second cumulant (second panel), excess kurtosis (third panel) and sextosis (fourth panel) for the spin-$\frac{1}{2}$ XXZ brickwork circuit at Free fermion point, $(\mathcal{J},\mathcal{J}^\prime)=(\frac{1}{4},0)$. Grey dashed line with circular markers indicate the results obtained from exact numerics from match-gate correspondence discussed here and red markers of different shades indicate the results obtained using the methodology presented in this work with different temporal bond dimensions, more redder indicates results obtained with larger bond dimensions (colorbar). In this free fermion limit, though accurate computation of local correlator needs larger bond dimensions, the full counting statistics results converge to exact numerical results with as small as $\chi=2^5$ bond dimension.}
\label{fig:bench_free}
\end{figure*}
\subsection{Dual Unitary Circuits}
At $\mathcal{J}=\frac{\pi}{4}$, the two site gates $\mathbb{U}_{n n+1}$ are dual unitary. In Dual unitary circuits composed of two-site gates, individual two site gates satisfy unitarity and unitality in the spatial direction~\cite{Bertini2019Exact,Gopalakrishnan2019Unitary,Bertini2025Exactly}, i.e.,
\begin{eqnarray}
&&\vcenter{\hbox{\includegraphics[width=.08\linewidth]{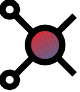}}}=\vcenter{\hbox{\includegraphics[width=.04\linewidth]{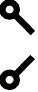}}} \hspace{.1\textwidth} \vcenter{\hbox{\includegraphics[width=.08\linewidth]{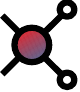}}}=\vcenter{\hbox{\includegraphics[width=.04\linewidth]{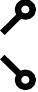}}}.
\end{eqnarray}
Using these identities iteratively, exact left and right influence matrices can be obtained as product states, i.e., matrix product states with bond dimension one. These are respectively given as,
\begin{eqnarray}
   && \vcenter{\hbox{\includegraphics[width=.2\linewidth]{im_l.pdf}}}=\vcenter{\hbox{\includegraphics[width=.028\linewidth]{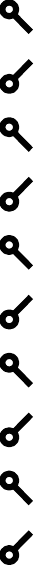}}}\hspace{.05\textwidth} \vcenter{\hbox{\includegraphics[width=.2\linewidth]{im_r.pdf}}}=\vcenter{\hbox{\includegraphics[width=.028\linewidth]{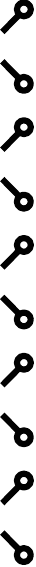}}}.
\end{eqnarray}
Using these exact expressions for the influence matrices gives the following expression for the moment generating function,
\begin{eqnarray}
    \mathcal{Z}(\lambda,n)&=&\vcenter{\hbox{\includegraphics[width=.33\linewidth]{z0lc.pdf}}}=\vcenter{\hbox{\includegraphics[width=.053\linewidth]{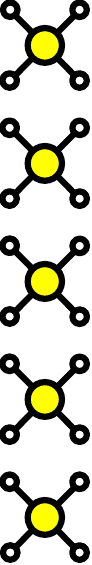}}}=\vcenter{\hbox{\includegraphics[width=.1\linewidth]{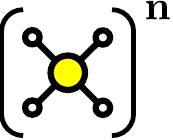}}},
\end{eqnarray}
which in algebraic form is given as,
\begin{eqnarray}
    \mathcal{Z}(\lambda,n) &=& \left(\frac{{\rm Tr}\left[e_{}^{i\lambda \left(\mathbb{M}_0 -\mathbb{M}_1\right)}\mathbb{U}_{01} e_{}^{-i\lambda \left(\mathbb{M}_0 -\mathbb{M}_1\right)}\mathbb{U}_{01}^\dagger\right]}{d^2}\right)_{}^n,\nonumber\\
\end{eqnarray}
which for the brickwork XXZ circuit at the dual unitary point ($\mathcal{J}=\frac{\pi}{4}$) is given by,
\begin{eqnarray}
\mathcal{Z}(\lambda,n) &=& \left[\frac{1+\cos\lambda}{2}\right]_{}^{n}.
\end{eqnarray}
It is interesting to note that, the full counting statistics at the dual unitary line is independent of $\mathcal{J}^\prime$.

Similarly, using the above expression for the exact influence matrix, the local magnetization correlator is given by $\langle Z_0(n) Z_0(0) \rangle=\delta_{n0}$, consistent with the well known fact that the correlation functions in dual unitary circuits are non-zero only on the light cone.

Although not shown, it is worth noting that our numerical simulation of the influence matrices does converge to the maximally mixed state as shown above and hence are consistent with the analytical results presented above.

\subsection{Numerical convergence analysis}
\begin{figure*}[htbp!]
\centering
\includegraphics[width=\linewidth]{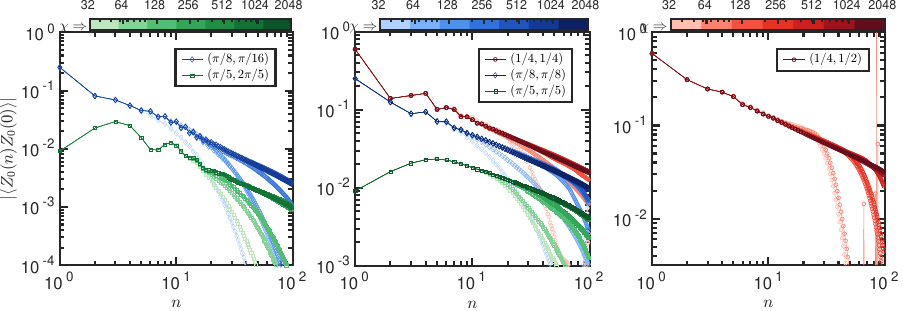}%
\caption{Local magnetization correlator ($\langle Z_0^{}(n) Z_0^{}(0)\rangle$) as a function of circuit depth ($n$) for varying values of bond dimensions ($\chi$) used for numerical simulation (colorbars).}
\label{fig:bench_szsz}
\end{figure*}
\begin{figure*}[htbp!]
\centering
\includegraphics[width=\linewidth]{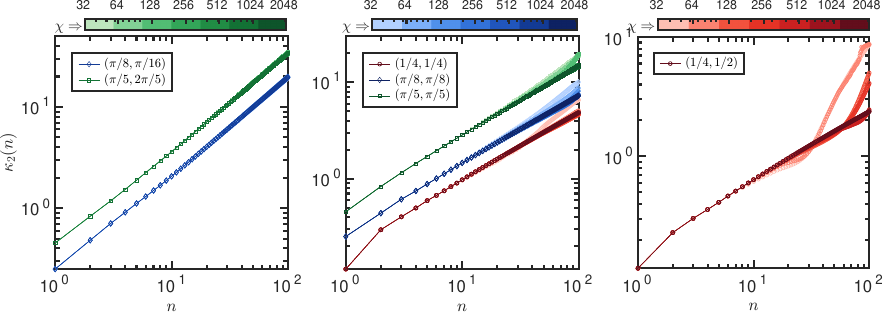}%
\caption{Second cumulant of magnetisation transferred across the interface ($\kappa_2(n)$) as a function of circuit depth ($n$) for varying values of bond dimensions ($\chi$) used for numerical simulation (colorbars).}
\label{fig:bench_k2}
\end{figure*}

Exact numerical results can be obtained with exact influence matrices, whose matrix product state representation needs exponential in the circuit depth. The numerical methodology presented in the main text approximates the influence matrices with a matrix product state with a fixed bond-dimension, this leads to errors in the observables presented. We do not aim to track the truncation error, as it is subtle with non-hermitian gate application, which is the case in transverse network contraction. Instead, the numerical results are converged by increasing bond dimension. 

\begin{figure*}[htbp!]
\centering
\includegraphics[width=\linewidth]{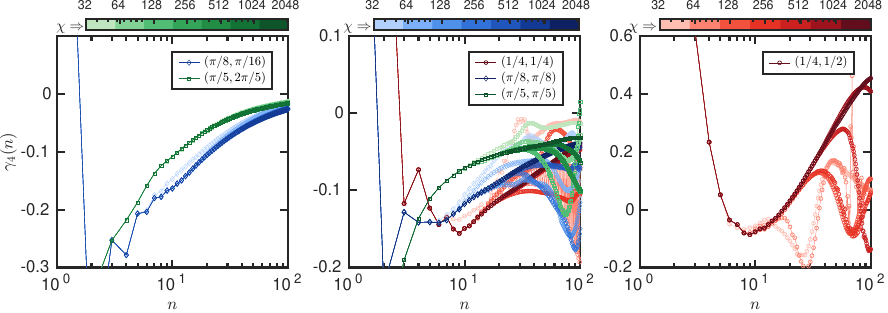}%
\caption{Excess kurtosis of magnetisation transferred across the interface ($\gamma_4(n)$) as a function of circuit depth ($n$) for varying values of bond dimensions ($\chi$) used for numerical simulation (colorbars).}
\label{fig:bench_g4}
\end{figure*}
\begin{figure*}[htbp!]
\centering
\includegraphics[width=\linewidth]{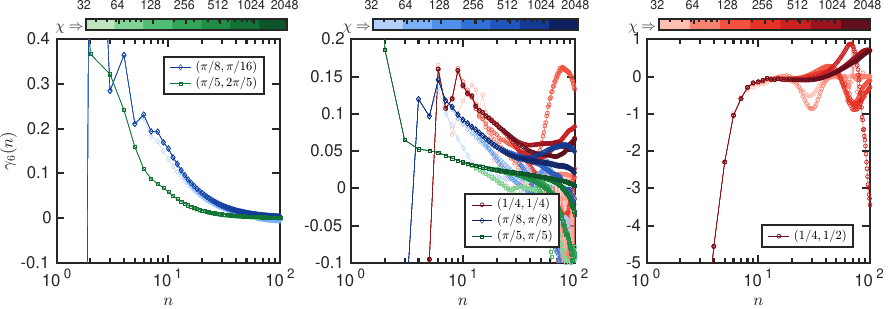}%
\caption{Sextosis of magnetisation transferred across the interface ($\gamma_6(n)$) as a function of circuit depth ($n$) for varying values of bond dimensions ($\chi$) used for numerical simulation (colorbars).}
\label{fig:bench_g6}
\end{figure*}

The local correlator, second cumulant, excess kurtosis and sextosis are presented in Figs. (\ref{fig:bench_szsz},\ref{fig:bench_k2},\ref{fig:bench_g4} \& \ref{fig:bench_g6}) for all the parameter regimes considered (except at $\mathcal{J}^\prime=0$) for increasing bond dimensions. A key point to note that, the results converge best in the ballistic regime, where the local correlator converges upto the circuit depths $n=10^2$ with bond dimension $\chi\sim2^8$, the cumulants converge must faster at bond dimension $\chi\sim2^6$. In the super-diffusive regime, the local correlator and second cumulant converge at $\chi\sim2^9$ for upto $n=10^2$. Converging higher cumulants requires larger bond dimensions, $\chi>2^{10}$. 
Diffusive regime is the hardest to obtain converged results. Though the local correlator and second cumulant converge at around $\chi\sim2^{10}$, the excess kurtosis and sextosis could only be converged reasonably for circuit depths $n\sim64$ with $\chi=2^{11}$. 

At $\mathcal{J}^\prime=0$, observables are also compared against numerically exact method exploiting the correspondence to match gate circuit and using free fermion methods in Fig. (\ref{fig:bench_free}). For this case, first few cumulants obtained with bond dimension as small as $\chi=2^5$. The local correlator ($\langle Z_0(n)Z_0(0)\rangle$) also converges with increasing bond dimensions upto the circuit depth $n=10^2$ considered in this work.

\bibliography{references}
\end{document}